\documentclass[preprint,showpacs,preprintnumbers,amsmath,amssymb]{revtex4}   
\usepackage{blindtext}
\usepackage{graphicx}
\usepackage{amssymb}
\usepackage{amsfonts}
\usepackage{amsmath}
\usepackage{textcomp}
\usepackage{xcolor}
\usepackage{ulem}
\usepackage{bm}
\usepackage{parskip,float,graphicx}
\usepackage{placeins}
\usepackage[caption = false]{subfig}
\usepackage[colorlinks=true,allcolors=blue]{hyperref}
\usepackage{physics}
\usepackage[separate-uncertainty=true,exponent-product=\!\cdot\!]{siunitx}

\begin{document}

\title{Spin-Hall devices: spin relaxation spatially separates current injection from Joule dissipation}
\author{Yanbo Qiao}
\affiliation{School of Physics, Peking University, Beijing 100871, China}
\affiliation{LSI, \'Ecole Polytechnique, CEA/DRF/IRAMIS, CNRS, Institut Polytechnique de Paris, 91120 Palaiseau, France}

\author{Sariah Al Saati} 
\affiliation{Centre de Physique Th\'eorique, \'Ecole Polytechnique, 91120 Palaiseau, France}

\author{J.-E. Wegrowe} \email{jean-eric.wegrowe@polytechnique.edu}
\affiliation{LSI, \'Ecole Polytechnique, CEA/DRF/IRAMIS, CNRS, Institut Polytechnique de Paris, 91120 Palaiseau, France}
\date{\today}

\begin{abstract}
The stationary state of a spin Hall bar connected to an external load circuit is investigated through a variational approach based on the principle of minimum power dissipation generalized to the two-spin-channel model. The self-consistent distributions of longitudinal and transverse current densities, alongside the corresponding spin and charge accumulations and dissipation power in the resistance, are derived. Surprisingly, it is shown that the Joule dissipation vanishes when the load resistance is placed at a sufficiently large distance compared with the spin-relaxation length. Such a highly non-trivial global stationary state appears as the most striking characteristic of the injection of pure spin current compared with more usual current injection.
\end{abstract}


\maketitle

\section{Introduction}
According to Rudolf Clausius, the distinction between equilibrium and non-equilibrium states is as fundamental to physics as the distinction between life and death  \cite{Clausius}. 
Dissipation produces entropy, thereby defining an arrow of time and making physical evolution irreversible, whereas equilibrium obeys detailed balance and exhibits no preferred direction of time.

In the framework of linear non-equilibrium thermodynamics, the stationary state of a constrained system is selected by the minimum entropy production. When the system is described solely by charge transport, this stationary state is uniquely determined by the electrical constraints. The situation is fundamentally different in the presence of internal degrees of freedom \cite{DeGroot}. Their relaxation introduces additional irreversible processes that enlarge the space of admissible stationary states and may profoundly modify the spatial distribution of entropy production.\\

\begin{figure} [ht]
   \begin{center}
   \includegraphics[height=7cm]{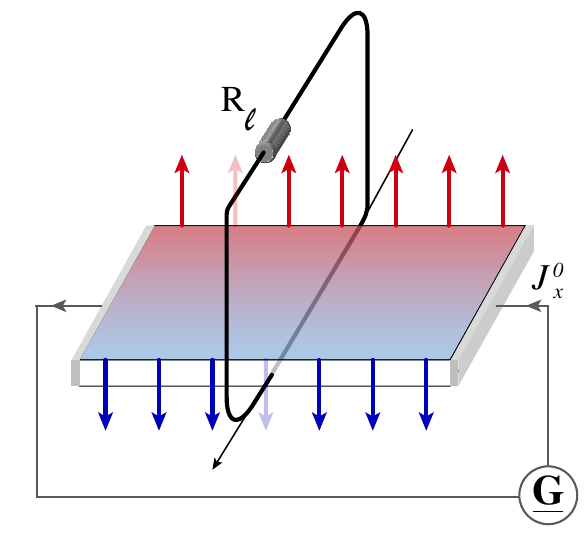}
      \includegraphics[height=7cm]{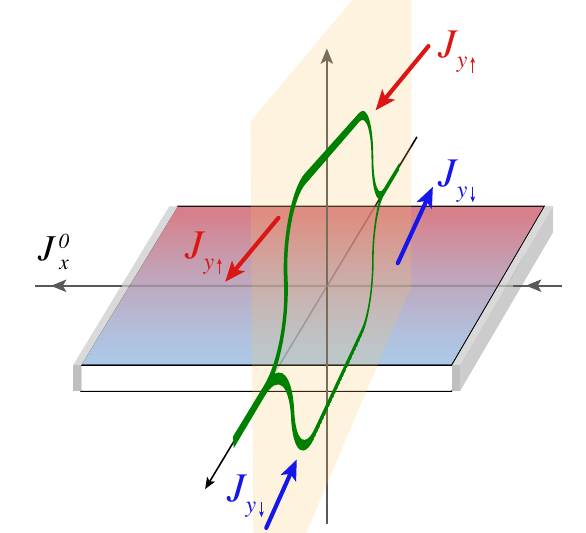}
   \end{center}
   \caption[example] 
   { \label{fig:Fig.2} 
Schematic representation of the spin-Hall bar connected to a load resistance $R_\ell$. Left: An electric generator imposes a longitudinal stationary current $J^0_x$ that produces spin accumulation at the edges. What is the value of the Joule dissipation in the load resistance? Right: Illustration of the injection of pure spin current within the two-spin-channel model. A pure spin current is injected from the spin-Hall bar into the load circuit (red arrows for the spin-up channel $J_{y,\uparrow}$ and blue arrows for the spin-down channel $J_{y,\downarrow}$). The vertical direction (profile in green) gives the amplitude of the spin-dependent currents in each channel. The spin polarization of the currents relaxes to zero over the spin-flip scattering length $\lambda_\text{sf}$.}
   \end{figure}

The spin Hall effect (SHE) \cite{Dyakonov,Hirsch,Zhang,Tse,Maekawa,Hoffmann,Bauer,Saslow,SHE,Taniguchi,JPhys17,PRBWeg} provides a particularly instructive realization of this situation. Spin-orbit coupling converts a part of the longitudinal charge current (injected by an electric generator) into a transverse pure spin current. In a spin-Hall device, the pure spin current can then be injected into a laterally connected load circuit, as sketched in Fig.~\ref{fig:Fig.2}. More precisely, a charge current generated by the spin-up channel ($\uparrow$) is injected from the edges of the spin-Hall bar into the lateral branch in one direction, while a charge current generated by the spin-down channel ($\downarrow$)---of identical amplitude---is injected in the other direction, $J_{y,\downarrow}=-J_{y,\uparrow}$.

In the spin-Hall device, the dissipation has two distinct origins: conventional Joule heating arising from the nonequilibrium charge distribution and spin-flip relaxation associated with the nonequilibrium spin accumulation. While the regime in which the load is located within one spin-relaxation length of the spin Hall bar is well known thanks to the studies performed in lateral spin valves \cite{Vila,Niimi RPB,Omori,Vila-2021,Casanova}, the opposite limit, where the load is placed at a large distance, has received little attention. It seems, however, that this configuration shows a remarkable distribution of entropy production. 

We investigate this regime within the framework of phenomenological non-equilibrium thermodynamics, for which the stationary state is defined by the minimum dissipation principle. The temperature is assumed constant, so that the dissipated power plays the role of the internal entropy production.  
The analysis is based on the two-spin-channel model, which has been intensively used in the context of giant magnetoresistance \cite{Johnson,Wyder,Fert,PRB2000,Jedema,Tulapurkar}, spin injection \cite{Schmidt,Smith,Jaffres}, and SHE \cite{Hirsch,Zhang,Tse,Maekawa,Hoffmann,Saslow,Taniguchi}. \\

Previous studies have shown theoretically \cite{JAP3,PRB-Sariah} and experimentally \cite{JAP4,JAP5} that a Hall bar (in the presence of an effective magnetic field but without explicit spin degrees of freedom) connected to a load circuit generates a ``standard'' \cite{Kirchhoff} thermodynamic distribution of entropy production: the current produced by the Hall effect induces Joule dissipation inside the load resistance.  In the present work, we generalize the variational approach by taking into account the spin degrees of freedom in the framework of the two conducting spin-channel model. We show that the relaxation of the spin degrees of freedom fundamentally changes the stationary state selected by the minimum dissipation principle. Entropy production becomes confined to the nanometric region connecting the Hall bar to the load, whereas electrical current injection occurs {\it with vanishing Joule dissipation} in the load circuit. The system thus provides an explicit example in which current injection and entropy production become spatially separated through the relaxation of internal degrees of freedom. In other words, the electric power injected transversely by the spin-Hall bar is {\it entirely transduced} into spin relaxation at the nanoscale. This property explains why---in the context of manipulating an adjacent ferromagnetic layer---the efficiency of lateral spin-orbit torque (SOT) \cite{SOT,SOTb} can be of the same order as that of direct spin-transfer torque (STT), despite the reduction of the lateral current injection associated with the small spin-Hall angle. \\

The paper is organized in six sections.
In section II below, we define the system under study, and we introduce the power functional for the bulk spin-Hall bar. The section III presents the stationary state {\it without load circuit}. 
Both the continuity equation and the boundary conditions are deduced from the minimization of the dissipation functional. The spatial distributions for the charge current, the spin current, the density of electric charges and the density of spins are derived, and the well-known results obtained in the pioneering works \cite{Dyakonov,Hirsch,Zhang,Tse,Maekawa,Hoffmann,Bauer,Saslow} are recovered.
Section IV presents the case of the stationary state including the load circuit placed at a large distance. The minimization of the functional that takes into account the bulk spin Hall bar and the load circuit is performed. The spatial distributions for the charge current, the spin current, the density of electric charges and the density of spins are deduced within an approximation based on a self-consistent perturbative approach. This calculation shows that the Joule dissipation inside the load resistance vanishes. 
In section V the question about the separation between current injection and Joule dissipation (i.e. entropy production) is discussed on the basis of the comparison with the usual and anomalous Hall effects, in which this separation does not exist. The conclusion summarizes the results derived in sections II to V and opens the discussion of the spin-orbit torque effect, which is of practical importance. 

\section{Thermodynamics of the spin-Hall bar}
Consider a homogeneous, non-magnetic conducting layer of length $L$, width $2\ell$ and thickness $d$, invariant under translation along the longitudinal $x$-axis. The system is characterized by an equilibrium background carrier density $2n_0$ and an electronic mobility $\eta$. Within the standard two-channel approximation, the charge carriers are decoupled into two independent populations denoted by the spin index $\sigma = \pm 1$ (corresponding to spin-up $\uparrow$ and spin-down $\downarrow$, respectively). The local carrier density for each subpopulation is denoted by $n_\sigma(y)$, while the local current distributions are described by the longitudinal components $J_{x\sigma}(y)$ and the transverse components $J_{y\sigma}(y)$.

Assuming spatial invariance along the transport axis ($\partial_x J_{i\sigma} = 0$, $i \in \{x,y\}$), the system is driven by a uniform longitudinal electrostatic field $E_x = -\partial_x V(x,y) = \text{const}$ sustained by the primary generator. Driven by this field, we define the uniform internal drift current density $J^0 \equiv q \eta n_0 E_x$.

Governed by the spin-dependent Ohm's law, the local transport properties can be expressed in these forms:
\begin{align}
    J_{x\sigma}(y) &= q \eta n_\sigma(y) \left[ E_x - \sigma \theta \partial_y \mu_\sigma(x, y) \right], \label{eq:transport_x} \\
    J_{y\sigma}(y) &= -q \eta n_\sigma(y) \left[ \partial_y \mu_\sigma(x, y) + \sigma \theta E_x \right], \label{eq:transport_y}
\end{align}
where $q >0$ is the charge of carriers, and $\theta$ denotes the effective spin Hall angle, assumed to be small ($\theta \ll 1$) in typical experimental regimes.

The local electrochemical potential $\mu_\sigma(x, y)$, expressed in units of volts (energy per unit charge), is linked to the carrier density via the state equation:
\begin{equation}
    \mu_\sigma(x, y) = \frac{kT}{q} \ln\left(\frac{n_\sigma(y)}{n_0}\right) + V(x, y) + \mu_\sigma^0.
    \label{eq: state_equation}
\end{equation}
Here, $T$ serves as the effective thermodynamic temperature of the carrier distribution (typically corresponding to the Fermi temperature $T_F$), $V(x,y)$ is the local electrostatic potential, and $\mu_\sigma^0$ is the equilibrium chemical potential baseline. Since the transverse density accumulation is invariant along $x$, the longitudinal gradient reduces to the electrostatic driving field, $-\partial_x \mu_\sigma = -\partial_x V = E_x$.

Substituting the transport relation Eq.~(\ref{eq:transport_y}) into the gradient of the spin accumulation potential $\Delta \mu(y) = \sum_\sigma \sigma \mu_\sigma(y)$ and integrating from the left boundary ($-\ell$) expresses $\Delta\mu$ as a functional of the transverse currents:
\begin{equation}
    \Delta \mu[J_{y\sigma}](y) = \Delta \mu(-\ell) - \int_{-\ell}^y \left[ \sum_{\sigma'} \frac{\sigma' J_{y\sigma'}(y')}{q \eta n_{\sigma'}(y')} + 2\theta E_x \right] \mathrm{d}y',
\end{equation}
where $\Delta \mu(-\ell)$ serves as an unconstrained integration constant.

In addition to the spin-dependent Ohm's law, the description of the electric transport has to be completed by the spin-flip process, occurring in the configuration space of the internal degrees of freedom. Now, the thermodynamic driving force is the spin accumulation potential $\Delta\mu(y)$, and its conjugate flux is the spin relaxation rate $\dot{\psi}(y)$. In the linear response regime, they are coupled via the effective Onsager coefficient $\tilde{\mathcal{L}}$ such that $\dot{\psi}(y) = \tilde{\mathcal{L}}\Delta\mu(y)$ \cite{Johnson,Wyder,Fert,PRB2000,Jedema,Schmidt,Smith,Jaffres,Tulapurkar}. Therefore, the reduced spin-flip power dissipation density is $\widetilde{P}_{\text{sf}} = \dot{\psi}\Delta\mu = \tilde{\mathcal{L}}(\Delta\mu)^2$. The coefficient $\tilde{\mathcal{L}}$ will be defined below in terms of the better-known spin-diffusion length $\lambda_\text{sf}$. 

To establish the variational scheme, we introduce the reduced total Joule dissipation functional $\widetilde{P}_{\text{tot}}$, which scales linearly with the total physical Joule power via $P_{\text{tot}} \equiv \frac{Ld}{q\eta(1+\theta^2)}\widetilde{P}_{\text{tot}}$. The relative dissipation in the external circuit is parameterized \cite{JAP3,PRB-Sariah} by the dimensionless control parameter $\alpha$, defined as
\begin{equation}
    \alpha \equiv \frac{2\ell}{R_\ell q n_0 \eta (1+\theta^2)} = \frac{R}{R_{\ell}}
\end{equation}
where $R_\ell$ is the resistance (per unit of $x$) of the lateral load connected across the boundaries and $R$ is the resistance (per unit of $x$) of the lateral part inside the Hall-bar, through which the pure spin-current is flowing.

The stationary state is governed by global requirements. We define the transverse spatial average for any local field $f(y)$ as $\langle f(y) \rangle \equiv \frac{1}{2\ell} \int_{-\ell}^{\ell} f(y) \, \mathrm{d}y$. 

First, the total charge conservation across the transverse cross-section requires the averaged carrier densities to satisfy:
\begin{equation}
    \sum_{\sigma} \langle n_\sigma(y) \rangle = 2n_0.
    \label{eq:constraint_n}
\end{equation}
Because spin-flip processes do not conserve the population of either channel separately, Eq.~(\ref{eq:constraint_n}) constrains only the total density. We therefore define the admissible density states by
\begin{equation*}
    \mathcal{A}_n \equiv \left\{(n_\uparrow,n_\downarrow): n_\sigma(y)>0,\quad \sum_\sigma\langle n_\sigma\rangle=2n_0\right\}.
\end{equation*}
Second, the galvanostatic condition dictating a constant total longitudinal charge current $2J_x^0$ entering from the longitudinal contacts requires:
\begin{equation}
    \sum_{\sigma} \langle J_{x\sigma}(y) \rangle = 2J_x^0.
    \label{eq:constraint_Jx}
\end{equation}

To evaluate the non-equilibrium driving sources developed at the interfaces, we define the reduced boundary driving force $A^\sigma$ for each channel. Obtained by averaging the local transverse flux related to Eq.~(\ref{eq:transport_y}), $A^\sigma$ measures the effective chemical potential drop across the sample width:
\begin{equation}
    A^\sigma[J;n] = - n_0 \left\langle \frac{\sigma\theta J_{x\sigma}(y) + J_{y\sigma}(y)}{n_\sigma(y)} \right\rangle.
    \label{eq:Asigma_def}
\end{equation}
With the reduced-power normalization introduced above, the reduced load dissipation is
\begin{equation}
    \widetilde P_{\rm load}[J;n]
    =
    \frac{2\alpha\ell}{n_0}
    \sum_{\sigma}
    \left(A^\sigma[J;n]\right)^2.
    \label{eq:Pload_spin_channels}
\end{equation}
For the isolated configuration,
$R_\ell\to\infty$ and hence $\alpha\to0$, so that
$\widetilde P_{\rm load}=0$.

The physical reduced bulk dissipation and the constrained functional are distinguished as follows. The current variations below are performed for an arbitrary admissible density profile $n\in\mathcal A_n$. Only the galvanostatic constraint requires a Lagrange multiplier $\lambda_x$:

\begin{equation}
\begin{aligned}
\widetilde{P}_{\text{bulk}}[J;n]
&= \int_{-\ell}^{\ell} \Bigg\{ \sum_{\sigma}
\frac{J_{x\sigma}^2(y)+J_{y\sigma}^2(y)}{n_\sigma(y)}
+\tilde{\mathcal{L}}\big(\Delta\mu[J_{y\sigma};n](y)\big)^2\Bigg\}\,\mathrm{d}y,\\
\widetilde{\mathcal F}_{\text{tot}}[J;n]
&=\widetilde{P}_{\text{bulk}}[J;n]+\widetilde{P}_{\text{load}}[J;n]
-\lambda_x\left(\sum_\sigma\langle J_{x\sigma}\rangle-2J_x^0\right),
\qquad n\in\mathcal A_n.
\label{Functional}
\end{aligned}
\end{equation}

Our model is valid under the following approximations: (i) the system maintains spatial invariance along the longitudinal axis ($\partial_x = 0$), neglecting finite-size corner effects at the external generator contacts; (ii) the transport relations and boundary driving forces are evaluated to the leading order in the perturbation parameters ($\theta \ll 1$ and $\delta n_\sigma \ll n_0$), thereby neglecting second-order density couplings in the conductivities; (iii) we assume the Thomas-Fermi screening length is strictly smaller than the spin diffusion length ($\tilde{\lambda}_D \ll \lambda_\text{sf}$), which decouples the boundary charge accumulation layer from the extended bulk spin diffusion profile; and (iv) the external circuit is modeled as an ideal, spin-independent {\color{red} resistance}, devoid of spin-polarization and parameterized solely by a scalar $\alpha$ (ratio of the resistances).

\section{Stationary state without load circuit}
We first consider an ideally isolated spin Hall bar, where the lateral boundaries are insulated. No external circuit is attached, yielding $\widetilde{P}_{\text{load}}^{(I)} = 0$. The open-circuit constrained functional is therefore $\widetilde{\mathcal F}_{\text{tot}}^{(I)}=\widetilde P_{\text{bulk}}-\lambda_x(\sum_\sigma\langle J_{x\sigma}\rangle-2J_x^0)$, at fixed $n\in\mathcal A_n$.

\subsection{State of minimum power dissipation}

In this subsection the variational variables are $J_{x\sigma}(y)$, $J_{y\sigma}(y)$, and $\Delta\mu(-\ell)$; no variation with respect to $n_\sigma(y)$ is performed. The density dependence is parametric in the flux minimization.

The integration constant $\Delta \mu(-\ell)$ acts as a free parameter. The stationary condition $\partial \widetilde{\mathcal F}_{\text{tot}}^{(I)} / \partial \Delta \mu(-\ell) = 0$ yields the global condition:
\begin{equation}
    \int_{-\ell}^{\ell} \Delta \mu(y) \, \mathrm{d}y = 0.
    \label{eq:zero_integral_condition}
\end{equation}

Minimizing with respect to the longitudinal current density at $n\in\mathcal A_n$, $\delta \widetilde{\mathcal F}_{\text{tot}}^{(I)} / \delta J_{x\sigma}=0$, gives $J_{x\sigma}(y)=\frac{\lambda_x}{4\ell}n_\sigma(y)$. The imposed current constraint Eq.~(\ref{eq:constraint_Jx}), together with the normalization already contained in $\mathcal A_n$, determines $\lambda_x$ and hence the local longitudinal current:
\begin{equation}
    J_{x\sigma}(y) = J_x^0 \frac{n_\sigma(y)}{n_0}.
    \label{eq:J_x}
\end{equation}
Taking the functional derivative with respect to the transverse current density at $n\in\mathcal A_n$, $\delta \widetilde{\mathcal F}_{\text{tot}}^{(I)} / \delta J_{y\sigma} = 0$, and exchanging the order of integration via Fubini's theorem isolates $\delta J_{y\sigma}$, yielding the integral equation:
\begin{equation}
    J_{y\sigma}(y) = \frac{\sigma \widetilde{\mathcal{L}}}{q \eta} \int_y^{\ell} \Delta \mu(y') \, \mathrm{d}y'.
    \label{eq:J_y}
\end{equation}

Differentiating Eq.~(\ref{eq:J_y}) with respect to $y$ recovers the {\it well-known continuity equation} governing spin-flip relaxation:
\begin{equation}
    \partial_y J_{y\sigma}(y) = - \frac{\sigma \widetilde{\mathcal{L}}}{q \eta} \Delta \mu(y).
    \label{Continuity}
\end{equation}

It is worth noting that this result is directly derived from of the minimization of the functional Eq.(\ref{Functional}) (instead of being a working hypothesis).

Summing this differential equation over the spin channels ($\sigma = \pm 1$) gives $\partial_y (J_{y\uparrow} + J_{y\downarrow}) = 0$, implying a constant total transverse charge current $J^Q_y\equiv J_{y\uparrow} + J_{y\downarrow} = C$. Evaluating Eq.~(\ref{eq:J_y}) at $y = \ell$ yields $J_{y\sigma}(\ell) = 0$. Evaluating at $y = -\ell$ yields $J_{y\sigma}(-\ell) = 0$ identically due to Eq.~(\ref{eq:zero_integral_condition}). Consequently, $C = 0$ and $J_{y\uparrow}(y) = -J_{y\downarrow}(y)$. The transverse flux is a pure spin current everywhere, emerging directly from the variational definition of the stationary state, without presuming local boundary conditions.

\subsection{Self-Consistent Resolution for currents and densities}

The variational relations derived above are now combined with the state equation, the transverse transport relation Eq.~(\ref{eq:transport_y}), and Poisson's equation. The continuity equation is already implied by Eq.~(\ref{eq:J_y}); together with the global normalization and boundary conditions, these relations determine the stationary current and density profiles as a coupled solution.
\subsubsection{Current}
The current fluxes are evaluated self-consistently with the carrier density $n_\sigma(y)$ and the non-equilibrium potential $\Delta\mu(y)$. Substituting the continuity equation {\color{red} Eq.(\ref{Continuity}) for $\sigma=1$} into the spatial derivative of the spin accumulation yields the differential equation for the transverse flux:
\begin{equation}
    \partial_y^2 J_{y\uparrow}(y) - \frac{1}{\lambda_\text{sf}^2} J_{y\uparrow}(y) = \frac{\theta J^0}{\lambda_\text{sf}^2},
\end{equation}
where $\lambda_\text{sf} \equiv l_{sf}/\sqrt{2(1+\theta^2)}$ is the effective spin diffusion length. This modified length scale emerges from the bulk spin-flip coupling coefficient $\widetilde{\mathcal{L}} = q\eta(1+\theta^2)\mathcal{L}$, where $\mathcal{L} = (2 q n_0 \eta) / (2 l_{sf}^2)$ is the standard Onsager coefficient for spin relaxation.

Imposing the derived boundary condition $J_{y\sigma}(\pm\ell)=0$ gives, within the approximation used to obtain the spin-diffusion equation, the transverse current for each channel:
\begin{equation} 
    J_{y\sigma}(y) = - \sigma \theta J^0 \left[ 1 - \frac{\cosh(y/\lambda_\text{sf})}{\cosh(\ell/\lambda_\text{sf})} \right].
    \label{eq: exact_Jy_isolated}
\end{equation}

Evaluating the cross-sectional longitudinal transport relation on this profile gives, within the same approximation,

\begin{equation}
    J_x^0 = J^0 (1 + \theta^2 f_\text{sf}),
\end{equation}
where we introduce the dimensionless spin-flip decay factor $f_\text{sf}$. This parameter emerges as the spatial average of the hyperbolic profile across the bar width:
\begin{equation}
    f_\text{sf} \equiv \left\langle \frac{\cosh(y/\lambda_\text{sf})}{\cosh(\ell/\lambda_\text{sf})} \right\rangle = \frac{\lambda_\text{sf}}{\ell} \tanh\left(\frac{\ell}{\lambda_\text{sf}}\right).
    \label{spin-flip}
\end{equation}

The solution recovers the two expected limiting behaviors. In the strong spin-flip limit, $\lambda_\text{sf}\ll\ell$, one has $f_\text{sf}\to0$. Away from the edge boundary layers of width $\sim\lambda_\text{sf}$, the transverse current approaches the Corbino-like bulk value,
$J_{y\sigma}\to-\sigma\theta J^0$, while $J_x^0\to J^0$. In the opposite limit of negligible spin-flip scattering, $\lambda_\text{sf}\gg\ell$, one has $f_\text{sf}\to1$; the transverse current vanishes throughout the bar, $J_{y\sigma}\to0$, and the longitudinal relation becomes $J_x^0\to J^0(1+\theta^2)$.

The transverse pure spin current is defined as the difference between the orthogonal spin fluxes, $J_y^{\text{Sp}}(y) = J_{y\uparrow}(y) - J_{y\downarrow}(y)$:

\begin{equation}
    J_y^{\text{Sp}}(y) = - 2 \theta J^0 \left[ 1 - \frac{\cosh(y/\lambda_\text{sf})}{\cosh(\ell/\lambda_\text{sf})} \right].
\end{equation}

\begin{figure}[htbp]
    \centering
    \includegraphics[width=0.78\textwidth]{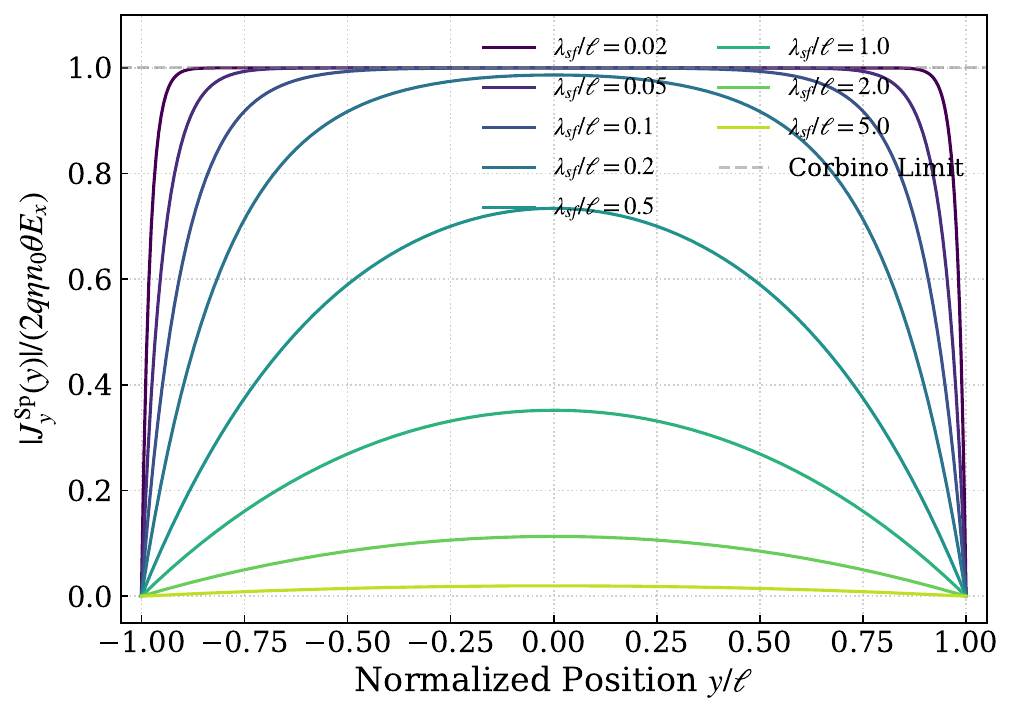}
    \caption{Normalized spatial profiles of the transverse spin current amplitude. The current is scaled by the ideal Corbino limit value $2 \theta J^0$. Curves correspond to varying characteristic length ratios $\lambda_{\text{sf}}/\ell$, with the dashed line indicating the Corbino limit.}
    \label{fig:jsp_profile}
\end{figure}

This profile satisfies the open-circuit boundary conditions $J_y^{\text{Sp}}(\pm \ell) = 0$, exhibiting exponential boundary layers of width $\sim \lambda_\text{sf}$ localized near the edges (see Fig.~\ref{fig:jsp_profile}). In the deep bulk region ($\ell - |y| \gg \lambda_\text{sf}$) under the strong spin-flip regime ($\ell/\lambda_\text{sf} \gg 1$), it recovers the Corbino limit $J_y^{\text{Sp, bulk}} \approx - 2 \theta J^0$.

\FloatBarrier

\subsubsection{Spin Accumulation}
For convenience, we define $\delta n_\sigma \equiv n_\sigma - n_0$, and the useful basis transformation $n^\text{Sp} \equiv n_\uparrow-n_\downarrow$ and $\delta n^Q \equiv \delta n_\uparrow + \delta n_\downarrow$, where $n^\text{Sp}$ denotes spin accumulation and $\delta n^Q$ denotes charge accumulation.

Applying the continuity equation Eq.(\ref{Continuity}) and substituting the parameter mapping for the effective spin diffusion length, the electrochemical potential splitting $\Delta\mu(y)$ is determined from the transverse flux:
\begin{equation}
    \Delta\mu(y) = - \frac{2 \theta J^0 \lambda_\text{sf}}{q \eta n_0} \frac{\sinh(y/\lambda_\text{sf})}{\cosh(\ell/\lambda_\text{sf})}.
\end{equation}

\begin{figure}[htbp]
    \centering
    \includegraphics[width=0.7\textwidth]{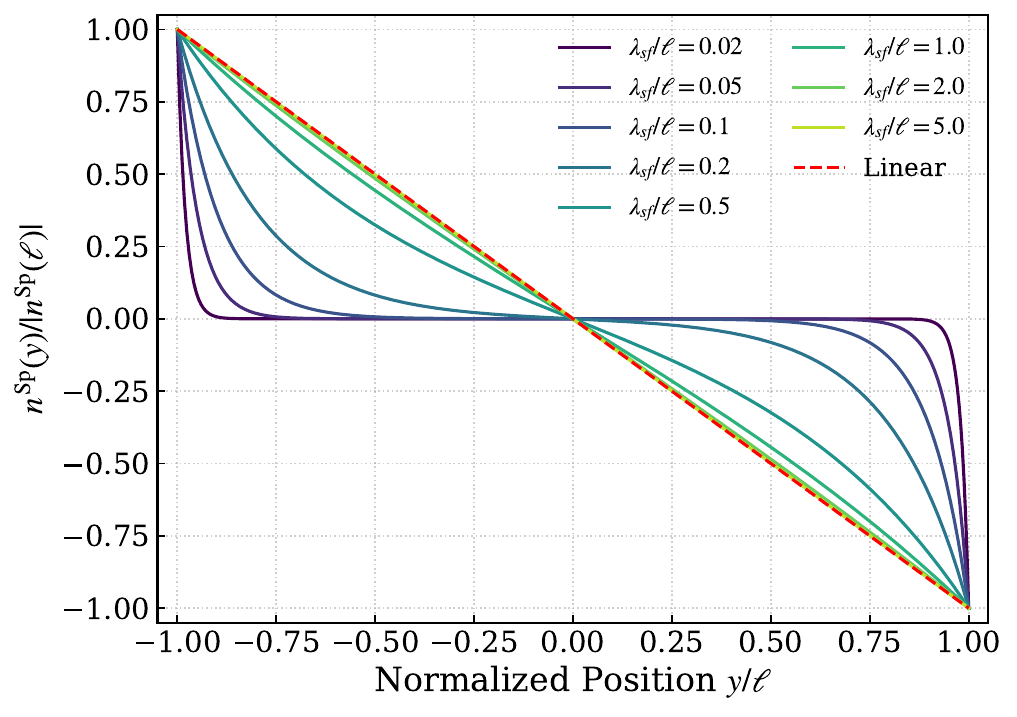}
    \caption{Normalized spatial profiles of the spin accumulation $n^{\text{Sp}}(y)$. Each profile is scaled by its absolute boundary value $|n^{\text{Sp}}(\ell)|$. The solid curves correspond to varying characteristic length ratios $\lambda_{\text{sf}}/\ell$. The red dashed line denotes the linear approximation asymptote corresponding to the limit of negligible spin-flip scattering.}
    \label{fig:nsp_profile}
\end{figure}

Under the approximation for the carrier density ($\delta n_\sigma \ll n_0$), the local spin accumulation scales proportionally to the localized chemical potential difference:
\begin{equation}
    n^{\text{Sp}}(y) \simeq \frac{q n_0}{kT} \Delta \mu(y) = - \frac{2 \theta J^0 \lambda_\text{sf}}{k T \eta} \frac{\sinh(y/\lambda_\text{sf})}{\cosh(\ell/\lambda_\text{sf})}.
    \label{eq:n_sp_approx}
\end{equation}
The transverse spin current decays over the scale $\lambda_\text{sf}$, yielding a hyperbolic spin accumulation profile across the diffusion length as expected for the SHE \cite{Dyakonov,Hirsch,Zhang,Tse,Maekawa,Hoffmann,Bauer,Saslow,SHE,Taniguchi}. As shown in Fig.~\ref{fig:nsp_profile}, this profile converges to the linear asymptote ($n^{\text{Sp}} \propto y$) in the weak spin-flip limit ($\ell/\lambda_\text{sf} \to 0$). In the strong spin-flip limit ($\ell/\lambda_\text{sf} \to \infty$), the boundary accumulation vanishes linearly (with respect to $\lambda_\text{sf}$).

\subsubsection{Longitudinal Spin Current}
Similar to the basis for transverse current, we also define the basis transformation for longitudinal current, $J^Q_x = J_{x\uparrow}+J_{x\downarrow}$ and $J^\text{Sp}_x = J_{x\uparrow}-J_{x\downarrow}$.

The longitudinal spin current $J_x^{\text{Sp}}(y)$ is obtained by taking the difference between the two orthogonal spin channels. Utilizing Eq.~(\ref{eq:J_x}), it evaluates to:
\begin{equation}
    J_x^{\text{Sp}}(y) = J_{x\uparrow}(y) - J_{x\downarrow}(y) = J_x^0 \frac{n^{\text{Sp}}(y)}{n_0}.
\end{equation}
Substituting the spin accumulation profile derived in Eq.~(\ref{eq:n_sp_approx}) yields:
\begin{equation}
    J_x^{\text{Sp}}(y) \simeq - \frac{2 \theta J_x^0 J^0 \lambda_\text{sf}}{k T \eta n_0} \frac{\sinh(y/\lambda_\text{sf})}{\cosh(\ell/\lambda_\text{sf})},
\end{equation}
which gives an antisymmetric profile. 
\subsubsection{Charge Accumulation and Screening}
We define
$\nu^Q\equiv\delta n^Q/n_0$ and
$\nu^{\rm Sp}\equiv n^{\rm Sp}/n_0$.
Summing the transverse transport relation
Eq.~(\ref{eq:transport_y}) over the two spin channels, using
$J_y^Q=0$, and substituting
$\partial_y\mu_\sigma=(kT/q)(\partial_y n_\sigma/n_\sigma)+\partial_yV$
gives
\begin{equation}
    \frac{kT}{qn_0}\partial_y\delta n^Q
    +(2+\nu^Q)\partial_yV
    =
    -\theta E_x\nu^{\rm Sp}.
    \label{eq:charge_gradient_unexpanded}
\end{equation}

The non-magnetic two-channel model is invariant under the
combined transformation $\theta\to-\theta$ and
$\sigma\rightarrow -\sigma$. Hence the spin response is odd in
$\theta$, whereas the induced charge and electrostatic responses are even.
Therefore we have 
\(
\nu^{\rm Sp}=\mathcal O(\theta),\) \(
\nu^Q=\mathcal O(\theta^2)\) and \(\partial_yV/E_x=\mathcal O(\theta^2).
\)\footnote{At this order, an explicit expansion of the spin-summed
logarithmic state equation gives
\(
\mu_\uparrow+\mu_\downarrow
=
(kT/q)\left[\nu^Q-\tfrac14(\nu^{\rm Sp})^2\right]
+2V+\mu_\uparrow^0+\mu_\downarrow^0
+\mathcal O(\theta^4).
\)
Thus, $\nu^Q$ and $(\nu^{\rm Sp})^2$ contribute at the same order.
The density-weighted transport sum leading to
Eq.~(\ref{eq:charge_gradient_unexpanded}) retains this quadratic
spin-density contribution automatically.}
Therefore $\nu^Q\partial_yV$ contributes only at fourth order,
and the first non-vanishing charge order is
\begin{equation}
    \frac{kT}{qn_0}\partial_y\delta n^Q
    +2\partial_yV
    =
    -\frac{\theta E_x}{n_0}n^{\rm Sp}.
    \label{eq:charge_gradient_consistent}
\end{equation}
Using $\partial_y^2V=-(q/\epsilon)\delta n^Q$ therefore yields
\begin{equation}
    \tilde\lambda_D^2\partial_y^2\delta n^Q-\delta n^Q
    =\mathcal S_Q(y),\qquad
    \mathcal S_Q(y)=-\frac{\epsilon\theta E_x}{2qn_0}\partial_y n^{\rm Sp},
    \label{eq:ode_charge}
\end{equation}
where $\tilde\lambda_D=\sqrt{\epsilon kT/(2q^2n_0)}$ is the effective Debye length and $\epsilon$ is the permittivity. For the open circuit,
\begin{equation}
    \mathcal S_Q(y)=\mathcal S_0\cosh\left(\frac{y}{\lambda_\text{sf}}\right),\qquad
    \mathcal S_0=\frac{\epsilon\theta^2(J^0)^2}
    {q^2\eta^2n_0^2kT\cosh(\ell/\lambda_\text{sf})}.
    \label{eq:SQ_open}
\end{equation}
The open-circuit charge problem is invariant under reflection
$y\to-y$, so $\delta n^Q(y)$ is even, which forces the solution to be
\begin{equation}
    \delta n^Q(y)
    =
    C_Q\cosh\left(\frac{y}{\tilde\lambda_D}\right)
    -
    \frac{\mathcal S_0\cosh(y/\lambda_\text{sf})}
    {1-(\tilde\lambda_D/\lambda_\text{sf})^2}.
    \label{eq:exact_charge_general}
\end{equation}

Following Ref.~\cite{PRBWeg}, we consider a globally neutral
Hall bar embedded in a symmetric electrostatic environment, yielding
    $\partial_yV(\pm\ell)=0$. Therefore, Equation~(\ref{eq:charge_gradient_consistent}) gives
\begin{equation}
    \partial_y\delta n^Q(\pm\ell)
    =-\frac{q\theta E_x}{kT}n^{\rm Sp}(\pm\ell)
    =\pm\frac{2\ell f_\text{sf}\theta^2(J^0)^2}
    {(kT\eta)^2n_0}.
    \label{eq:boundary_gradient_Q_open}
\end{equation}
This boundary condition fixes
\begin{equation}
    C_Q=\frac{\lambda_\text{sf}\mathcal S_0\sinh(\ell/\lambda_\text{sf})}
    {\tilde\lambda_D[1-(\tilde\lambda_D/\lambda_\text{sf})^2]
    \sinh(\ell/\tilde\lambda_D)}.
    \label{eq:CQ_constant}
\end{equation}
Although $C_Q$ is exponentially small for $\tilde\lambda_D\ll\ell$, its homogeneous term supplies the narrow Debye edge layer.

The resulting solution confirms the order assignment used above:
\(\nu^{\mathrm{Sp}}=\mathcal O(\theta)\) and
\(\nu^Q=\mathcal O(\theta^2)\). Consequently, charge-density feedback
modifies the leading spin profiles only at order \(\theta^3\), as quantified
below.

\begin{figure}[htbp]
    \centering
    \includegraphics[width=0.7\textwidth]{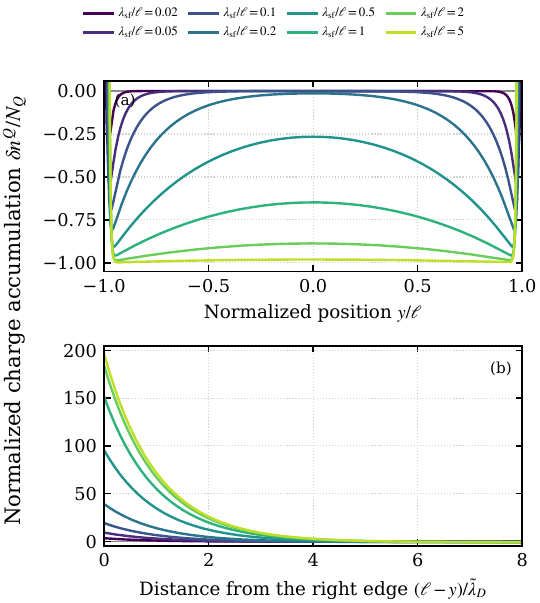}
    \caption{Corrected charge-accumulation profiles normalized by
$N_Q=\epsilon\theta^2E_x^2/kT$, for
$\tilde\lambda_D/\ell=0.005$ and varying
$\lambda_{\rm sf}/\ell$. Panel (a) displays the negative component on the
spin-diffusion scale. Panel (b) resolves the positive Debye layer near the
right boundary using the scaled distance
$(\ell-y)/\tilde\lambda_D$; the left boundary layer is identical by
reflection symmetry.}
    \label{fig:nQ_macro}
\end{figure}

\section{Spin-Hall bar connected to a lateral load placed at a large distance}
We now resolve the steady state for a spin Hall bar connected to an external load at a large distance with respect to the spin-diffusion length $\lambda_\text{sf}$. The load is characterized by the ratio of resistances $\alpha = R/R_{\ell} > 0$.

To decouple the charge and spin transport at the boundaries, we transform the reduced driving sources into the charge and spin basis: $A^Q \equiv A^\uparrow + A^\downarrow$ and $A^{\text{Sp}} \equiv A^\uparrow - A^\downarrow$. The external load dissipation functional (defined in Eq.~(\ref{eq:Pload_spin_channels})) can thus be diagonalized into two independent macroscopic thermodynamic channels:
\begin{equation}
\widetilde{P}_{\text{load}}^{(II)} = \alpha \frac{\ell}{n_0} \left[ (A^Q)^2 + (A^{\text{Sp}})^2 \right].
\end{equation}

\subsection{State of minimum dissipation}

As in the open-circuit case, the functional derivatives are taken with respect to the current fields and $\Delta\mu(-\ell)$ while holding $n_\sigma$ fixed in those derivatives.

The variation with respect to the boundary potential parameter $\Delta\mu(-\ell)$ remains independent of $A^\sigma$. Thus, it enforces the same spin-neutrality condition derived for the open-circuit limit:
\begin{equation}
    \int_{-\ell}^{\ell} \Delta \mu(y) \, \mathrm{d}y = 0.
\end{equation}

Minimizing the constrained functional with respect to the longitudinal current, $\delta \widetilde{\mathcal F}_{\text{tot}}^{(II)} / \delta J_{x\sigma}=0$ at $n\in\mathcal A_n$, yields:
\begin{equation}
    J_{x\sigma}(y) = \frac{\lambda_x}{4\ell} n_\sigma(y) + \alpha \sigma \theta A^\sigma.
\end{equation}
Summing over the spin channels and integrating across the transverse width, the imposed current constraint Eq.~(\ref{eq:constraint_Jx}) and the density normalization contained in $\mathcal A_n$ determine $\lambda_x$. Substituting $\lambda_x$ back isolates the longitudinal current:
\begin{equation}
    J_{x\sigma}(y) = J_x^0 \frac{n_\sigma(y)}{n_0} + \alpha \theta \left[ \sigma A^\sigma - \frac{n_\sigma(y)}{2n_0} A^\text{Sp} \right].
    \label{eq:Jx_load}
\end{equation}

Taking the functional derivative with respect to the transverse current density, $\delta \widetilde{\mathcal F}_{\text{tot}}^{(II)} / \delta J_{y\sigma}=0$ at $n\in\mathcal A_n$, yields the integral equation:
\begin{equation}
    J_{y\sigma}(y) = \frac{\sigma \widetilde{\mathcal{L}}}{q \eta} \int_y^{\ell} \Delta \mu(y') \, \mathrm{d}y' + \alpha A^\sigma.
    \label{eq:Jy_load}
\end{equation}
Evaluating Eq.~(\ref{eq:Jy_load}) at $y = \ell$ defines the non-zero transverse boundary extraction imposed by the load:
\begin{equation}
    J_{y\sigma}(\ell) = \alpha A^\sigma.
    \label{eq:J_ysigma b.c.}
\end{equation}
Differentiating Eq.~(\ref{eq:Jy_load}) with respect to $y$ recovers the bulk continuity equation $\partial_y J_{y\sigma}(y) = - (\sigma \widetilde{\mathcal{L}} / q \eta) \Delta \mu(y)$. This confirms that the external load modifies the transverse boundary conditions while preserving the continuity equation.

\subsection{Self-Consistent Resolution at leading order}

To derive the spin diffusion equation, we first simplify the transport equations using an approximation. We apply the zeroth-order approximation for the carrier density, $n_\sigma(y) \simeq n_0$, within the denominators of the transport relations. This approximation neglects second-order couplings between the local currents and the non-equilibrium density variations ($\delta n_\sigma \ll n_0$), while preserving the first-order gradients in $\Delta \mu(y)$.

Under this approximation, substituting the charge and spin basis into the transport equations yields the gradient of the spin accumulation potential:
\begin{equation}
    \partial_y \Delta \mu(y) = - \frac{J_y^{\text{Sp}}(y)}{q \eta n_0} - 2\theta E_x.
\end{equation}

Taking the derivative of $J_y^{\text{Sp}}(y)$ and substituting the spin diffusion length definition yields the differential equation:
\begin{equation}
    \partial_y^2 J_y^{\text{Sp}}(y) - \frac{1}{\lambda_\text{sf}^2} J_y^{\text{Sp}}(y) = \frac{2 \theta J^0}{\lambda_\text{sf}^2}.
    \label{eq: J_Sp ode}
\end{equation}

We solve Eq.~(\ref{eq: J_Sp ode}) with the boundary condition $J_y^{\text{Sp}}(\pm \ell) = \alpha A^{\text{Sp}}$ derived from Eq.~(\ref{eq:J_ysigma b.c.}). The resulting transverse spin current is:
\begin{equation}
    J_y^{\text{Sp}}(y) = - 2 \theta J^0 \left[ 1 - \frac{\cosh(y/\lambda_\text{sf})}{\cosh(\ell/\lambda_\text{sf})} \right] + \alpha A^{\text{Sp}} \frac{\cosh(y/\lambda_\text{sf})}{\cosh(\ell/\lambda_\text{sf})}.
    \label{eq:JySp_load}
\end{equation}

Using the relation $J_{y\sigma} = \frac{1}{2}(J_y^Q + \sigma J_y^{\text{Sp}})$, the transverse flux for each independent spin channel is:
\begin{equation}
\begin{split}
    J_{y\sigma}(y) &= - \sigma \theta J^0 \left[ 1 - \frac{\cosh(y/\lambda_\text{sf})}{\cosh(\ell/\lambda_\text{sf})} \right] \\
    &\quad + \frac{\alpha}{2} \left[ A^Q + \sigma A^{\text{Sp}} \frac{\cosh(y/\lambda_\text{sf})}{\cosh(\ell/\lambda_\text{sf})} \right].
\end{split}
\label{eq:Jy_unified}
\end{equation}

Equations (\ref{eq:Jx_load}) and (\ref{eq:Jy_unified}) parameterize the current densities in terms of the boundary sources $A^Q$ and $A^{\text{Sp}}$. Substituting these back into the definition of the boundary driving force $A^\sigma$ (Eq.~(\ref{eq:Asigma_def})), we obtain the self-consistent equations for $A^\sigma$. Under the zeroth-order approximation ($n_\sigma \simeq n_0$), the basis sources are evaluated via their spatial averages:
\begin{align}
    A^Q &= -  \big\langle \theta J_x^\text{Sp} + J_y^Q \big\rangle, \label{eq:AQ_integral} \\
    A^{\text{Sp}} &= -  \big\langle \theta J_x^Q + J_y^{\text{Sp}} \big\rangle. \label{eq:ASp_integral}
\end{align}

Substituting the current profiles into Eq.~(\ref{eq:AQ_integral}) yields the self-consistent equation for the symmetric charge source:
\begin{equation}
    A^Q = - \alpha (1 + \theta^2) A^Q.
\end{equation}
Given $\alpha>0$, this condition enforces $A^Q=0$. Consequently, no charge leaks into the external load ($J_y^Q=0$), and the total longitudinal charge current density reduces at this order to $J_x^Q=2J_x^0$ (equivalently, each channel has the baseline $J_x^0$).

At the retained order, Eq.~(\ref{eq:ASp_integral}) and the averaged spin-current profile instead close the variational model directly:
\begin{equation}
    \begin{aligned}
    A^{\rm Sp}&=-2\theta J_x^0-\langle J_y^{\rm Sp}\rangle
    +\mathcal O(\theta^3J_x^0),\\
    \langle J_y^{\rm Sp}\rangle
    &=-2\theta J_x^0(1-f_\text{sf})
    +\alpha f_\text{sf}A^{\rm Sp}
    +\mathcal O(\theta^3J_x^0).
    \end{aligned}
    \label{eq:ASp_variational_closure}
\end{equation}
Consequently, we get:
\begin{equation}
    A^{\rm Sp}=-\frac{2\theta J_x^0f_\text{sf}}
    {1+\alpha f_\text{sf}}+\mathcal O(\theta^3J_x^0).
    \label{eq:ASp_solution}
\end{equation}

\subsection{Transport Profiles and Spin Accumulation}
To write the leading expressions compactly, we define
\begin{equation}
    A^* \equiv \frac{\theta J_x^0}{1+\alpha f_\text{sf}}.
\end{equation}
Substituting the source $A^{\text{Sp}}$ back into the transport profiles yields the transverse currents:
\begin{align}
    J_{y\sigma}(y) &= - \sigma A^* \left[ 1 + \alpha f_\text{sf} - \frac{\cosh(y/\lambda_\text{sf})}{\cosh(\ell/\lambda_\text{sf})} \right]
    +\mathcal O(\theta^3J_x^0), \label{eq:Jy_load_sigma} \\
    J_y^{\text{Sp}}(y) &= - 2 A^* \left[ 1 + \alpha f_\text{sf} - \frac{\cosh(y/\lambda_\text{sf})}{\cosh(\ell/\lambda_\text{sf})} \right]
    +\mathcal O(\theta^3J_x^0). \label{eq:Jy_load_sp}
\end{align}

The above spin-current profile $J_y^{\text{Sp}}(y)$ is plotted in Fig.5 for several values of $\alpha$ (at $\lambda_\text{sf}/\ell = 0.2$), and in Fig.6 for several values of $\lambda_\text{sf}/\ell$ (at $\alpha = 1.0$).

\begin{figure}[htbp]
    \centering
    \includegraphics[width=0.78\textwidth]{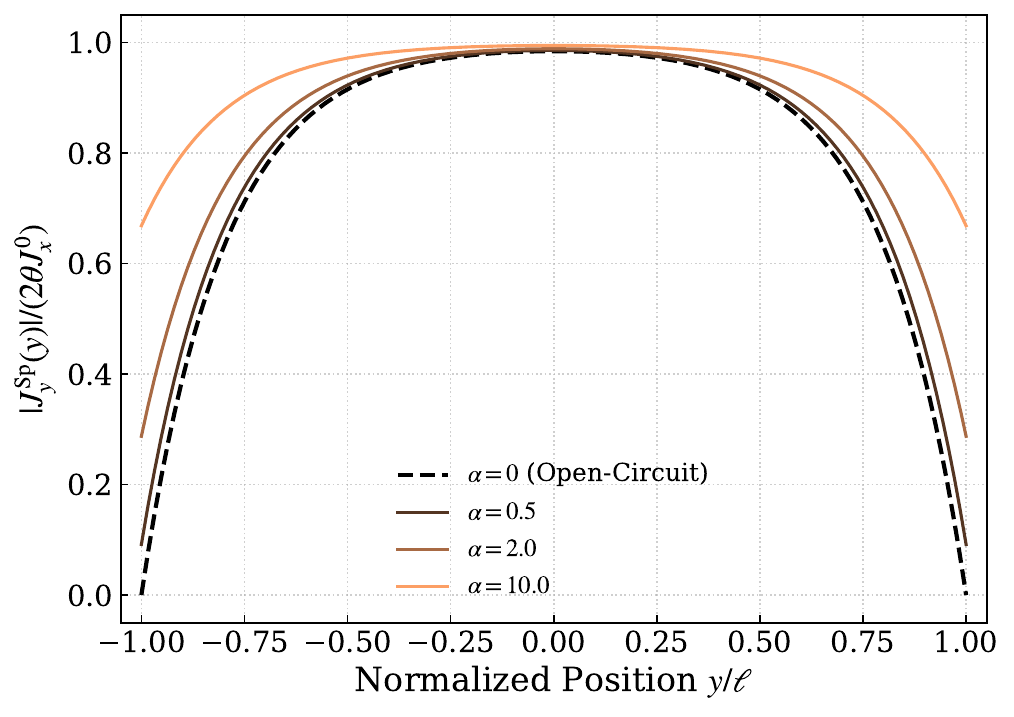}
    \caption{Spatial distribution of the transverse spin current $J_y^{\text{Sp}}(y)$ normalized by $2\theta J_x^0$. Parameters are $\lambda_\text{sf}/\ell = 0.2$ and $\theta = 0.1$. The dashed line ($\alpha=0$) corresponds to the ideal open-circuit limit.}
    \label{fig: JySp_alpha}
\end{figure}

\begin{figure}[htbp]
    \centering
    \includegraphics[width=0.78\textwidth]{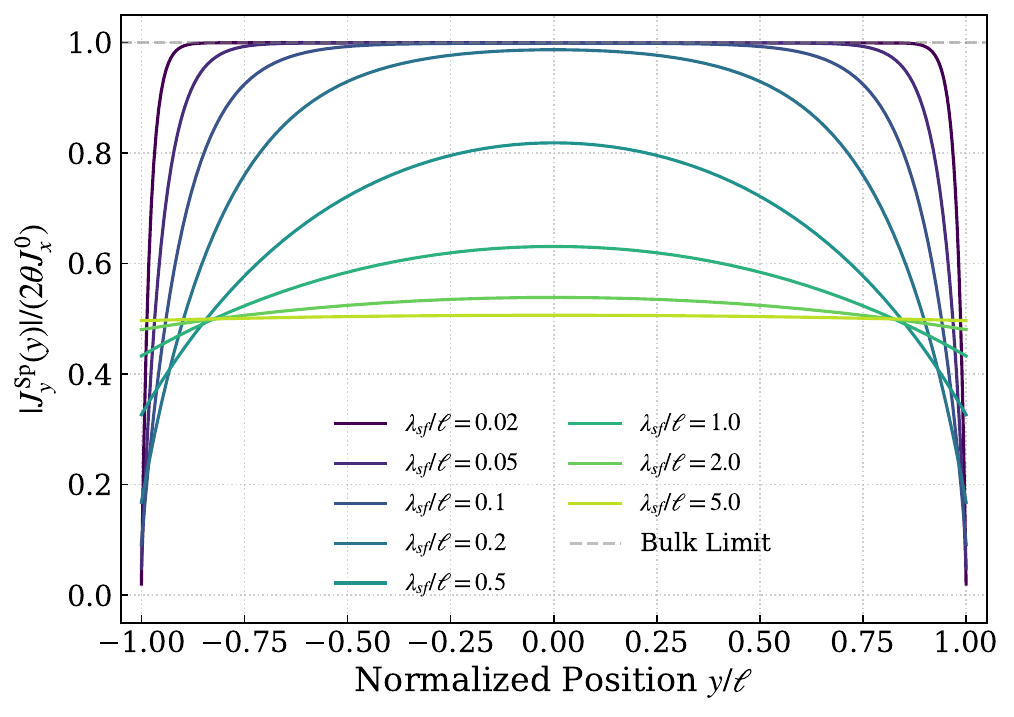}
    \caption{Spatial distribution of the transverse spin current $J_y^{\text{Sp}}(y)$ normalized by $2\theta J_x^0$. Parameters are $\alpha = 1.0$ and $\theta = 0.1$.}
    \label{fig: JySp_lambda}
\end{figure}

The spatial profile is governed by three distinct contributions: the ideal Corbino bulk drift, the shift induced by the external load, and the spatial relaxation due to spin-flip scattering. 

Evaluating the transverse flux at the boundaries gives
$J_y^{\rm Sp}(\pm\ell)=-2A^*\alpha f_\text{sf}+\mathcal O(\theta^3J_x^0)$, which quantifies the spin-current leakage to the external load and vanishes in the open-circuit limit ($\alpha\to0$).

Using the relation $\partial_y J_y^{\text{Sp}} = -(2\widetilde{\mathcal{L}}/q\eta)\Delta \mu$ and $n^{\text{Sp}} \simeq (q n_0 / k T) \Delta\mu$, the spatial profile of the spin accumulation is determined:
\begin{equation}
    n^{\text{Sp}}(y) = - \frac{2 \lambda_\text{sf}}{k T \eta} A^* \frac{\sinh(y/\lambda_\text{sf})}{\cosh(\ell/\lambda_\text{sf})}
    +\mathcal O\!\left(\frac{\theta^3J_x^0\lambda_\text{sf}}{kT\eta}\right).
\end{equation}

\begin{figure}[htbp]
    \centering
    \includegraphics[width=0.78\textwidth]{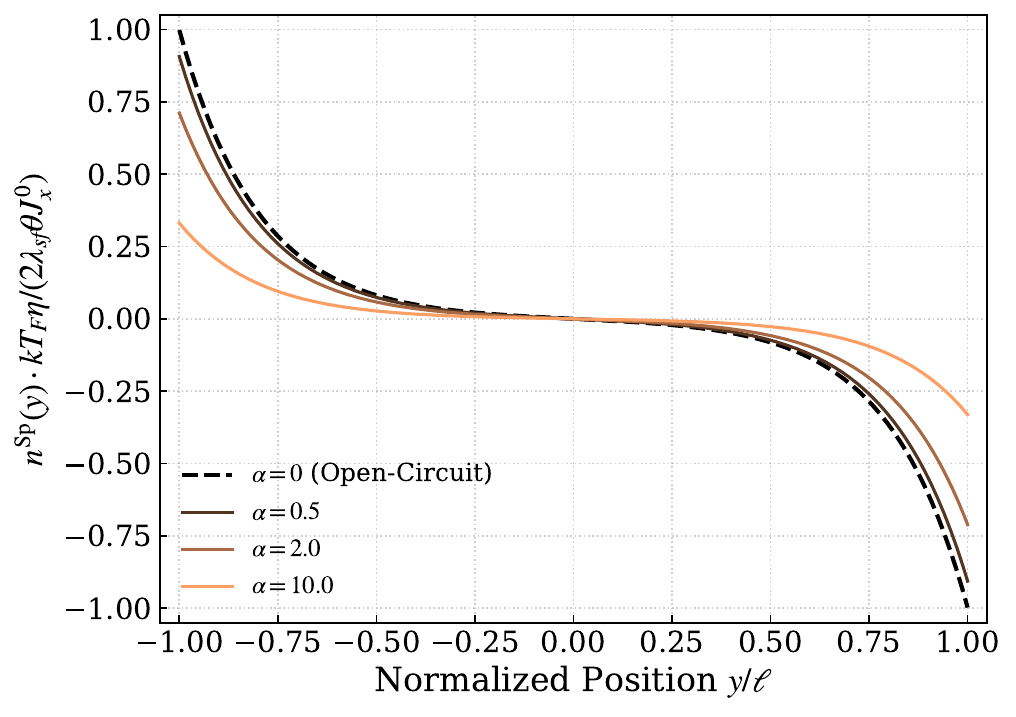}
    \caption{Spatial distribution of the spin accumulation $n^{\text{Sp}}(y)$ normalized by $2 \lambda_\text{sf} \theta J_x^0 / (k T \eta)$. Parameters are $\lambda_\text{sf}/\ell = 0.2$ and $\theta = 0.1$.}
    \label{fig: nSp_alpha}
\end{figure}

\begin{figure}[htbp]
    \centering
    \includegraphics[width=0.78\textwidth]{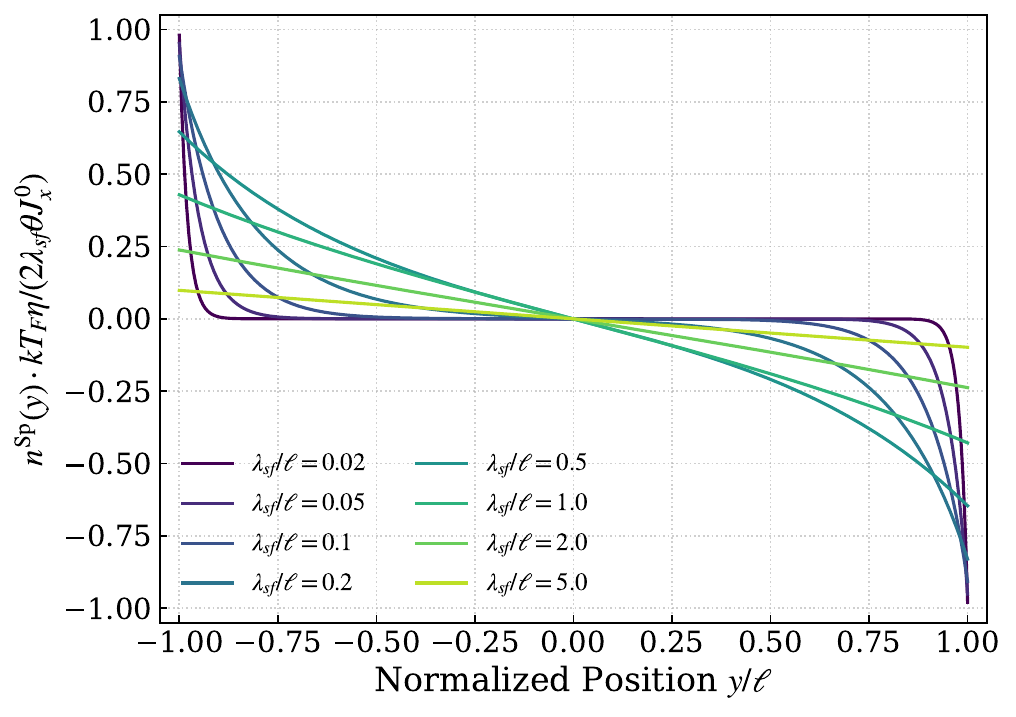}
    \caption{Spatial distribution of the spin accumulation $n^{\text{Sp}}(y)$ normalized by $2 \lambda_\text{sf} \theta J_x^0 / (k T \eta)$. Parameters are $\alpha = 1.0$ and $\theta = 0.1$.}
    \label{fig: nSp_lambda}
\end{figure}

\FloatBarrier
\subsection{Charge Accumulation under Load}
Since the stationary loaded state satisfies
$A^Q=0$ and hence $J_y^Q=0$, the charge equation
Eq.~(\ref{eq:ode_charge}) remains applicable. Substituting the loaded spin
accumulation gives the source
\begin{equation}
    \mathcal S_Q(y)
    =
    \mathcal S_1
    \cosh\left(\frac{y}{\lambda_\text{sf}}\right),
    \qquad
    \mathcal S_1
    =
    \frac{\epsilon(A^*)^2(1+\alpha f_\text{sf})}
    {q^2n_0^2kT\eta^2\cosh(\ell/\lambda_\text{sf})}.
    \label{eq:SQ_load}
\end{equation}

At fixed longitudinal drive, $A^*=\mathcal O(\theta)$, so
Eq.~(\ref{eq:SQ_load}) gives the first non-vanishing charge response.
For the same symmetric electrostatic environment as in the open-circuit
case, $\partial_yV(\pm\ell)=0$, and
Eq.~(\ref{eq:charge_gradient_consistent}) gives
\begin{equation}
    \partial_y\delta n^Q(\pm\ell)
    =
    \pm
    \frac{2\ell f_\text{sf}(1+\alpha f_\text{sf})(A^*)^2}
    {(kT\eta)^2n_0}.
    \label{eq:boundary_gradient_Q_load}
\end{equation}

Together with reflection symmetry, this boundary condition
determines the charge profile
\begin{equation}
    \delta n^Q(y)
    =
    C_Q\cosh\left(\frac{y}{\tilde\lambda_D}\right)
    -
    \frac{\mathcal S_1\cosh(y/\lambda_\text{sf})}
    {1-(\tilde\lambda_D/\lambda_\text{sf})^2},
    \label{eq:exact_charge_load}
\end{equation}
where
\begin{equation}
    C_Q
    =
    \frac{\lambda_\text{sf}\mathcal S_1
    \sinh(\ell/\lambda_\text{sf})}
    {\tilde\lambda_D
    [1-(\tilde\lambda_D/\lambda_\text{sf})^2]
    \sinh(\ell/\tilde\lambda_D)}.
    \label{eq:CQ_load}
\end{equation}

With this value of $C_Q$, the profile satisfies both
Eq.~(\ref{eq:boundary_gradient_Q_load}) and the global neutrality condition
$\int_{-\ell}^{\ell}\delta n^Q(y)\,\mathrm{d}y=0$. It reduces to the
open-circuit result in the limit $\alpha\to0$, through the retained order.

\begin{figure}[htbp]
    \centering
    \includegraphics[width=0.78\textwidth]{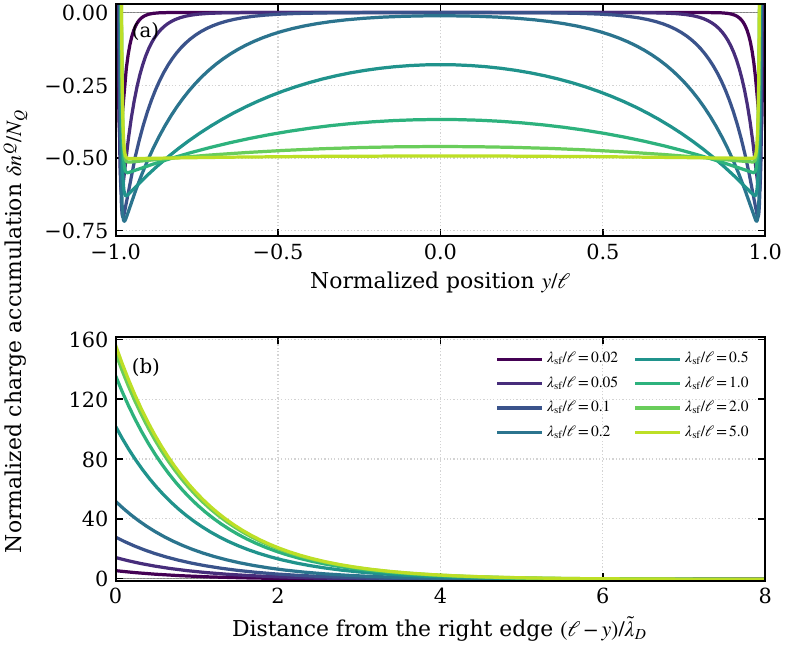}
    \caption{Charge-accumulation profiles under load, normalized
    by
    $N_Q=\epsilon\theta^2(J_x^0)^2/(q^2\eta^2n_0^2kT)$, for
    $\tilde\lambda_D/\ell=0.00317$, $\alpha=1$, and the values of
    $\lambda_{\rm sf}/\ell$ indicated in the legend. Panel (a) shows the profiles on the
    spin-diffusion scale. Panel (b) resolves the right Debye boundary layer
    as a function of $(\ell-y)/\tilde\lambda_D$; the left boundary layer
    follows by reflection symmetry.}
    \label{fig:nQ_spatial_profile}
\end{figure}

\subsection{Robustness of the leading-order approximation
within the variational framework}

To verify the leading-order approximation used in deriving the boundary
sources, let
$\nu_\sigma(y)=\delta n_\sigma(y)/n_0$ and introduce the dimensionless
driving parameter
$\chi=qE_x\lambda_{\rm sf}/(kT)\ll1$.%
\footnote{For typical metallic thin films, a current density
$J_x\sim1.6\times10^{11}\ {\rm A/m^2}$ and a resistivity
$\rho\sim10^{-7}\ \Omega\,{\rm m}$ give
$E_x\sim1.6\times10^4\ {\rm V/m}$. Taking
$\lambda_{\rm sf}\sim10\ {\rm nm}$ and an effective Fermi-energy scale
$kT/q\simeq4\ {\rm V}$ gives $\chi\sim4\times10^{-5}$. An effective
Debye length $\tilde\lambda_D\sim0.1\ {\rm nm}$ further gives
$(\tilde\lambda_D/\lambda_{\rm sf})^2\sim10^{-4}$.}
The profiles obtained above imply
$\nu^{\rm Sp}=\mathcal O(\theta\chi)$ and
\[
    \left\langle|\nu^Q|\right\rangle
    =
    \mathcal O\!\left[
        \theta^2\chi^2
        \left(\frac{\tilde\lambda_D}{\lambda_{\rm sf}}\right)^2
    \right].
\]
Although the local charge density is enhanced inside the narrow Debye
layers, their width is of order $\tilde\lambda_D$, so this enhancement
does not change the spatially averaged scaling. Hence
$\langle|\nu^Q|\rangle\ll
\langle|\nu^{\rm Sp}|\rangle\ll1$ for the parameter range considered here.

We now substitute
Eq.~(\ref{eq:Jx_load}) into
Eq.~(\ref{eq:Asigma_def}) and expand
$(1+\nu_\sigma)^{-1}$. In the charge combination, the only
inhomogeneous term at linear density order is proportional to
$\langle J_y^{\rm Sp}\nu^{\rm Sp}\rangle$. Reflection symmetry makes
$J_y^{\rm Sp}$ even and $\nu^{\rm Sp}$ odd, so this average vanishes.
Together with $J_y^Q=\alpha A^Q$ and global electroneutrality
$\langle\nu^Q\rangle=0$, the charge-source equation remains homogeneous
and its regular symmetric solution is $A^Q=J_y^Q=0$ through the retained
order.

For the spin combination, the same expansion gives
\begin{equation}
    A^{\rm Sp}
    =
    -2\theta J_x^0
    -\left\langle J_y^{\rm Sp}\right\rangle
    +\mathcal R_n,
    \qquad
    \mathcal R_n
    =
    \frac12\left\langle J_y^{\rm Sp}\nu^Q\right\rangle
    -\frac14\left\langle
        J_y^{\rm Sp}(\nu^{\rm Sp})^2
    \right\rangle
    +\cdots .
\end{equation}
Since
$\|J_y^{\rm Sp}\|_\infty=\mathcal O(\theta J_x^0)$, where \(\|f\|_\infty
\equiv
\max_{|y|\leq\ell}|f(y)|,\) the two displayed
corrections scale respectively as
\[
    \mathcal O\!\left[
        \theta^3\chi^2
        \left(\frac{\tilde\lambda_D}{\lambda_{\rm sf}}\right)^2
        J_x^0
    \right]
    \quad\text{and}\quad
    \mathcal O(\theta^3\chi^2J_x^0).
\]
Therefore
$\mathcal R_n=\mathcal O(\theta^3\chi^2J_x^0)$, which is already
contained in the $\mathcal O(\theta^3J_x^0)$ remainder of
Eq.~(\ref{eq:ASp_solution}). The density feedback consequently does not
modify the leading spin-current profiles or the leading load-power
result.

\subsection{Power Dissipation in the Load}
To assess how the two spin channels modify the normal Hall response, we calculate the Joule dissipation rate in the external load. Since $A^Q = 0$, the load functional reduces to $\widetilde{P}_{\text{load}} = \alpha \frac{\ell}{n_0} (A^{\text{Sp}})^2$. Substituting the transverse spin driving source (Eq.~\ref{eq:ASp_solution}) yields:
\begin{equation}
    \widetilde{P}_{\text{load}}
    =\frac{4\ell\alpha f_\text{sf}^2}{n_0}(A^*)^2
    +\mathcal O\!\left(\frac{\theta^4\ell(J_x^0)^2}{n_0}\right).
\end{equation}

To map this generalized dissipation to the physical dissipated power $P_{\text{load}}$ expressed in watts, we apply the relation $P_{\text{load}} = \frac{Ld}{q\eta(1+\theta^2)} \widetilde{P}_{\text{load}}$. We normalize this output against the open-circuit longitudinal Joule heating of the bulk sample, $P_0 =  (J_x^0)^2 \frac{4\ell  L d}{q \eta n_0}$. The normalized power, retained through its first non-zero order, is
\begin{equation}
    \frac{P_{\text{load}}}{P_0}
    =\frac{\theta^2\alpha f_\text{sf}^2}
    {(1+\alpha f_\text{sf})^2}+\mathcal O(\theta^4).
    \label{eq:P_load_absolute}
\end{equation}

This relation gives the load-power dependence on the control parameter $\alpha$ and the spin Hall angle $\theta$. This is one of the main results of this work, as it is directly accessible to measurement. We observe that this Joule power vanishes in both the open-circuit ($\alpha \to 0$) and short-circuit ($\alpha \to \infty$) limits.

The formal maximum of Eq.~(\ref{eq:P_load_absolute}) occurs at
$\alpha_{\rm opt}=f_\text{sf}^{-1}[1+\mathcal O(\theta^2)]$, corresponding to
$R_{\ell,\rm opt}/R=f_\text{sf}[1+\mathcal O(\theta^2)]$.  In the
strong-relaxation regime this is a low-load-resistance matching condition,
close to but distinct from the strict short-circuit limit
$\alpha\to\infty$.  Since a generic device is not assumed to track this
matching condition, we focus on the full load-power curve rather than on its
formal optimized value.

\begin{figure}[htbp]
    \centering
    \includegraphics[width=0.78\textwidth]{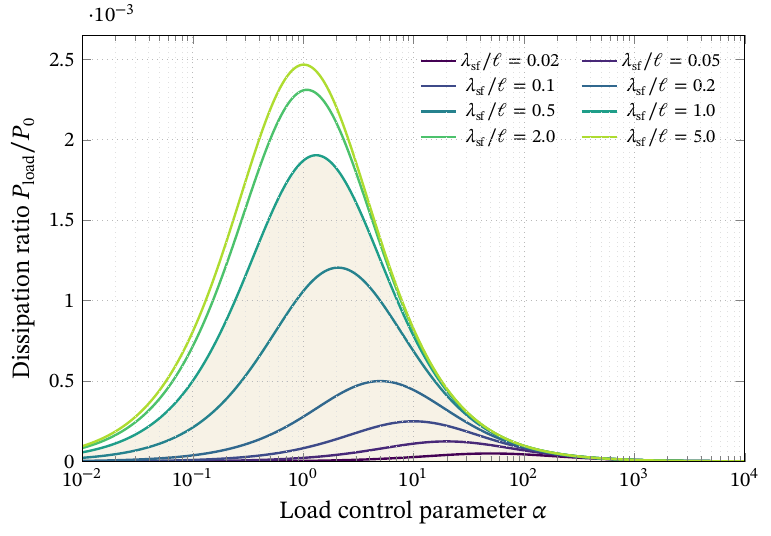}
    \caption{Normalized load dissipation
    $P_{\text{load}}/P_0$ as a function of $\alpha=R/R_\ell$ for several
    spin-diffusion ratios $\lambda_\text{sf}/\ell$, with $\theta=0.1$.
    The shaded envelope below the $\lambda_\text{sf}/\ell=1$ curve
    represents the strong-spin-relaxation regime
    $\lambda_\text{sf}/\ell<1$ relevant to usual spin-Hall bars. Curves
    with $\lambda_\text{sf}/\ell\geq1$ illustrate the crossover toward
    negligible spin relaxation.}
    \label{fig: P_load_P0}
\end{figure}

For any fixed $\alpha$, the $\lambda_\text{sf}/\ell=1$ curve
bounds from above all curves with $\lambda_\text{sf}/\ell<1$.
Thus, as shown in Fig.~\ref{fig: P_load_P0}, the shaded envelope
represents the strong-spin-relaxation regime
$\lambda_\text{sf}/\ell<1$. The comparison with the
weak-relaxation AHE limit is given in the next section.

\section{Comparison between SHE and AHE}

The load-power expression derived for the SHE enables a comparison with
the AHE results reported in Refs.~\cite{JAP4,JAP5}. Within the present
reduced description, the relevant distinction is that the transverse spin
current in the SHE is attenuated by an internal spin-relaxation channel,
whereas no analogous relaxing transverse degree of freedom is included in
the AHE model.

In the formal limit of negligible spin relaxation across the bar,
$\lambda_\text{sf}/\ell\to\infty$, the spin-flip factor approaches unity,
$f_\text{sf}\to1$. At the level of the normalized load-power law,
substituting $f_\text{sf}=1$ into Eq.~(\ref{eq:P_load_absolute})
reproduces, through the retained order, the AHE expression obtained in
Refs.~\cite{JAP3,PRB-Sariah}:
\begin{equation}
    \frac{P_{\text{AHE}}}{P_0}
    =
    \frac{\theta^2\alpha}{(1+\alpha)^2}
    +\mathcal O(\theta^4),
    \label{eq:P_AHE}
\end{equation}
where $\theta$ is understood as the corresponding Hall angle in each
model.

For comparison, the AHE expression is maximal at the matching condition
$\alpha=1$, giving
\begin{equation}
    \frac{P_{\text{AHE}}^{\text{max}}}{P_0}
    =
    \frac{\theta^2}{4}
    +\mathcal O(\theta^4).
\end{equation}
For the SHE, we instead consider a prescribed external load. In
the strong-relaxation regime, provided that
$\alpha f_\text{sf}\ll1$, Eq.~(\ref{eq:P_load_absolute}) yields
\begin{equation}
    \frac{P_{\text{SHE}}}{P_0}
    =
    \theta^2\alpha f_\text{sf}^2
    +\mathcal O(\theta^2\alpha^2f_\text{sf}^3)
    +\mathcal O(\theta^4)
    \propto
    \left(\frac{\lambda_\text{sf}}{\ell}\right)^2.
\end{equation}
Thus the quadratic length scaling applies to a prescribed
load only under the stated condition. For general $\alpha$, the full
expression Eq.~(\ref{eq:P_load_absolute}) should be used; when
$\alpha f_\text{sf}\gg1$, it approaches the short-circuit tail
$P_{\text{SHE}}/P_0\simeq\theta^2/\alpha$.
\section{Conclusion}

We have investigated the stationary state of a spin-Hall device composed of a spin-Hall bar connected to a load resistance (see Fig.~\ref{fig:Fig.2}). The spin-Hall bar injects a pure spin current: a charge current is injected from the edges of the spin-Hall bar into the lateral branch in one direction by the spin-up channel ($J_{y,\uparrow}$), while a charge current of identical amplitude is injected in the opposite direction by the spin-down channel ($J_{y,\downarrow}=-J_{y,\uparrow}$). This configuration has been investigated in the framework of studies about lateral spin valves \cite{Vila,Niimi RPB,Omori,Vila-2021,Casanova}. 

 In the study proposed above, the question was to derive the properties of the injected pure spin current {\it in a load resistance placed at a macroscopic distance from the edges}. The two spin currents, flowing in opposite directions, relax within the spin-diffusion length, but what is the power associated with the resulting flow? In usual Hall devices (without explicit spin degrees of freedom), in ferromagnets or in altermagnets (for the anomalous Hall effect), a systematic Joule dissipation is measured in the load resistance \cite{JAP4,JAP5}, which is approximately of the order of the square of the Hall angle $\theta^2$.

 Here we demonstrate that, for a prescribed load satisfying $\alpha f_\text{sf}\ll1$, spin-flip scattering suppresses the stationary Joule dissipation by the small factor $\theta^2(\lambda_\text{sf}/\ell)^2$, where typically $\lambda_\text{sf}/\ell\sim10^{-3}$ in realistic devices. In other words, {\it spin relaxation spatially separates current injection from Joule dissipation}. 
 
 The demonstration is based on the functional minimization of the total power dissipated in the system under the constraints imposed by the electric generator. The current density, chemical potential, and charge density distributions are defined throughout the Hall bar and the load resistance. The well-known results about SHE \cite{Dyakonov,Hirsch,Zhang,Tse,Maekawa,Hoffmann,Bauer,Saslow} were first recovered in the case of the open circuit (infinite load resistance). The calculation is then generalized with the load branch connected to the Hall bar. 

Furthermore, these results help explain why the spin-orbit torque effect can remain efficient despite a relatively small spin-Hall angle $\theta$: the lateral current injection is entirely converted into spin-relaxation processes:  the coupling with the ferromagnetic degrees of freedom of an adjacent layer is then optimal. 


\end{document}